\documentclass[aps,prd,twocolumn,amsmath,amssymb,floatfix,nofootinbib,superscriptaddress]{revtex4-1}
\usepackage{graphicx}
\usepackage{dcolumn}
\usepackage{bm}
\usepackage[usenames, dvipsnames]{color}
\usepackage[normalem]{ulem}
\usepackage{subfig}
\usepackage{sidecap}

\usepackage[colorlinks,bookmarks]{hyperref}
\definecolor{linkblue}{rgb}{0,0,0.8}
\definecolor{linkgreen}{rgb}{0,0.5,0}

\hypersetup{pdfpagemode=UseNone, pdfstartview=FitH, linkcolor=linkblue, %
            citecolor=linkgreen, urlcolor=linkblue}

\graphicspath{{Figures/}}

\newcommand{\kavli}{Kavli Institute for Cosmology, Madingley Road, Cambridge, UK, CB3 0HA}
\newcommand{\damtp}{DAMTP, Centre for Mathematical Sciences, Wilberforce Road, Cambridge, UK, CB3 0WA}
 
\newcommand{\VSI}{Van Swinderen Institute for Particle Physics and Gravity,\\ University of Groningen,
Nijenborgh 4, 9747 AG Groningen, The Netherlands}

\begin{document}

\title{Initial conditions of the universe: A sign of the sine mode}

\author{Darsh Kodwani}
\email{darsh.kodwani@physics.ox.ac.uk}
\affiliation{University of Oxford, Department of Physics, Denys Wilkinson Building, Keble Road, Oxford, OX1 3RH, UK.}

\author{P. Daniel Meerburg}
\email{pdm@ast.cam.uk}
\affiliation{\kavli}
\affiliation{\damtp}
\affiliation{\VSI}

\author{Ue-Li Pen}
\email{pen@cita.utoronto.ca}
\affiliation{Canadian Institute of Theoretical Astrophysics, 60 St George St, Toronto, ON M5S 3H8, Canada.}
\affiliation{Canadian Institute for Advanced Research, CIFAR program in Gravitation and Cosmology.}
\affiliation{Dunlap Institute for Astronomy \& Astrophysics, University of Toronto, AB 120-50 St. George Street, Toronto, ON M5S 3H4, Canada.}
\affiliation{Perimeter Institute of Theoretical Physics, 31 Caroline Street North, Waterloo, ON N2L 2Y5, Canada.}

\author{Xin Wang}
\email{wangxin35@mail.sysu.edu.cn}
\affiliation{School of Physics and Astronomy, Sun Yat-sen University, 2 Daxue Road, Zhuhai, China}
\affiliation{Canadian Institute of Theoretical Astrophysics, 60 St George St, Toronto, ON M5S 3H8, Canada.}

\begin{abstract}

In the standard big bang model the universe starts in a radiation dominated era, where the gravitational perturbations are described by second order differential equations, which will generally have two orthogonal set of solutions.
One is the so called {\it growing(cosine)} mode and the other is the {\it decaying(sine)} mode, where the nomenclature is derived from their behaviour on super-horizon(sub-horizon) scales. 
In most cosmological analysis it is assumed that only the growing mode is a viable solution, because on very large scales and early times the decaying solution shows singular behaviour and the amplitude of the mode is also highly suppressed in many inflationary models. 
However, physically interesting models do exist that would allow for decaying solutions, such as models in which the Universe today originates from a bounce.  Without singling out a specific model, an interesting and valid question is if a decaying mode
can actually result in a sensible cosmology, and withstand current precision cosmological constraints. 
The decaying mode is qualitatively different to the growing mode of adiabatic perturbations as it evolves with time on \emph{super-horizon} scales.
The time dependence of this mode on super-horizon scales is analysed in both the synchronous gauge and the Newtonian gauge to understand the true gauge invariant behaviour of these modes.
We then provide a gauge invariant procedure of normalising this mode on sub-horizon scales .
Then we explore constraints on the amplitude of this mode on scales between $k \sim 10^{-5}$ Mpc$^{-1}$ and $k \sim 10^{-1}$ Mpc$^{-1}$ using the temperature and polarization anisotropies from the cosmic microwave background, by computing the Fisher information.
Binning the primordial power non-parametrically into 100 bins, we find that the decaying modes are constrained at comparable variance as the growing modes on scales smaller than the horizon today using temperature anisotropies. Adding polrisation data makes the decaying mode more constrained. The decaying mode amplitude is thus constrained by $\sim 1/l$ of the growing mode.  On super-horizon scales, the growing mode is poorly constrained, while the decaying mode cannot substantially exceed the scale-invariant amplitude.  This interpretation differs substantially from the past literature, where the constraints were quoted in gauge-dependent variables, and resulted in illusionary tight super-horizon decaying mode constraints.
The results presented here can generally be used to non-parametrically constrain any model of the early universe. 

\end{abstract}

\maketitle


\section{Introduction and historical context}

Our current understanding of the Universe builds upon a widely accepted standard big bang model, in which the Universe starts out in a hot and dense radiation dominated phase. Precise initial conditions and an explanation of the homogeneity and isotropy of the large scale Universe are required to match current observations.
An epoch of cosmological inflation has been the most widely accepted extension to the standard big bang model that could potentially resolve these issues. Most importantly, it provides a natural way to generate small perturbations in the metric and densities of particles that manifest themselves as the temperature and polarization anisotropies in the cosmic microwave background (CMB) and density fluctuations that eventually grow into the large scale structure, which have been studied extensively over the last few decades \cite{0067-0049-208-2-20, Planck1,  refId0}.
The simplest models of inflation predict Gaussian adiabatic initial conditions for the radiation dominated era. 
However these are not the only possible initial conditions. .  
 
After neutrino decoupling at around $z\sim 10^9$, the universe contains baryons, photons, dark matter and neutrinos.
Each of these species has an equation governing its perturbations which are described by second order partial differential equations. 
In total there are 8 possible solutions for the densities of the particles that can exist in the early universe; two corresponding to each species \cite{Ma:1995ey, Bucher:1999re, Carrilho:2018mqy, Bucher:2000cd, Bucher:2000kb, Bucher:2000hy}. 
These solutions fall into two general classes, {\it adiabatic} or curvature and {\it entropy} or isocurvature fluctuations. 
The adiabatic solutions are defined as the solutions of the differential equations in which the relative number densities of all the particle species are the same. 
These are known as curvature perturbations as they correspond to an overall shift in the curvature of spacelike surfaces. 
On the contrary, the isocurvature perturbations correspond to solutions where the fractional number density of the particle species is not the same on spacelike surfaces.
Thus isocurvature perturbations are defined between any two species. For example, there can be a relative difference in the densities or velocities of the baryons and cold dark matter.
Canonically the isocurvature is defined as the fractional difference in particle species to the photon density.
In general there can be isocurvature between any of the particle species and therefore the most general initial conditions are given by a set of five possible linear combinations of modes: Adiabatic modes, CDM isocurvature, Baryon isocurvature, Neutrino density isocurvature and Neutrino velocity isocurvature \cite{Ma:1995ey, Bucher:2000cd, Bucher:2000kb, Bucher:2000hy}. 
There have been many attempts to constrain the amplitude of these general set of initial conditions and most studies show that the amplitude of isocurvature fluctuations must be much smaller than the amplitude of adiabatic fluctuations \cite{Moodley:2004nz, Beltran:2004uv, Lazarides:2004we, Beltran:2005xd}.
There is a further class of isocurvature known as {\it compensated isocurvature} in which there are isocurvature fluctuations due to both baryons and dark matter. 
This type of isocurvature has been shown to be more compatible with current observations \cite{Grin:2011tf, Grin:2013uya, Munoz:2015fdv}. 

In this study we do not consider isocurvature modes, instead we analyse the {\it structure} of adiabatic modes. 
Since the differential equations that govern all perturbations are second order differential equations, even for the adiabatic solution, there are two possible modes. 
One is called the {\it decaying} mode and the other is the more familiar {\it growing} mode. 
These names are motivated by the early time, super-horizon behaviour of these modes, as the decaying mode has a decaying behaviour whereas the growing mode remains constant.
The amplitude of these modes is usually set initially during a pre-radiation dominated era. Since the perturbation solution is a linear combination of each of these modes, \emph{both of these modes will be sourced by any pre-radiation phase that gives rise to adiabatic initial conditions}.

\begin{figure}[h!]
\includegraphics[scale=0.25]{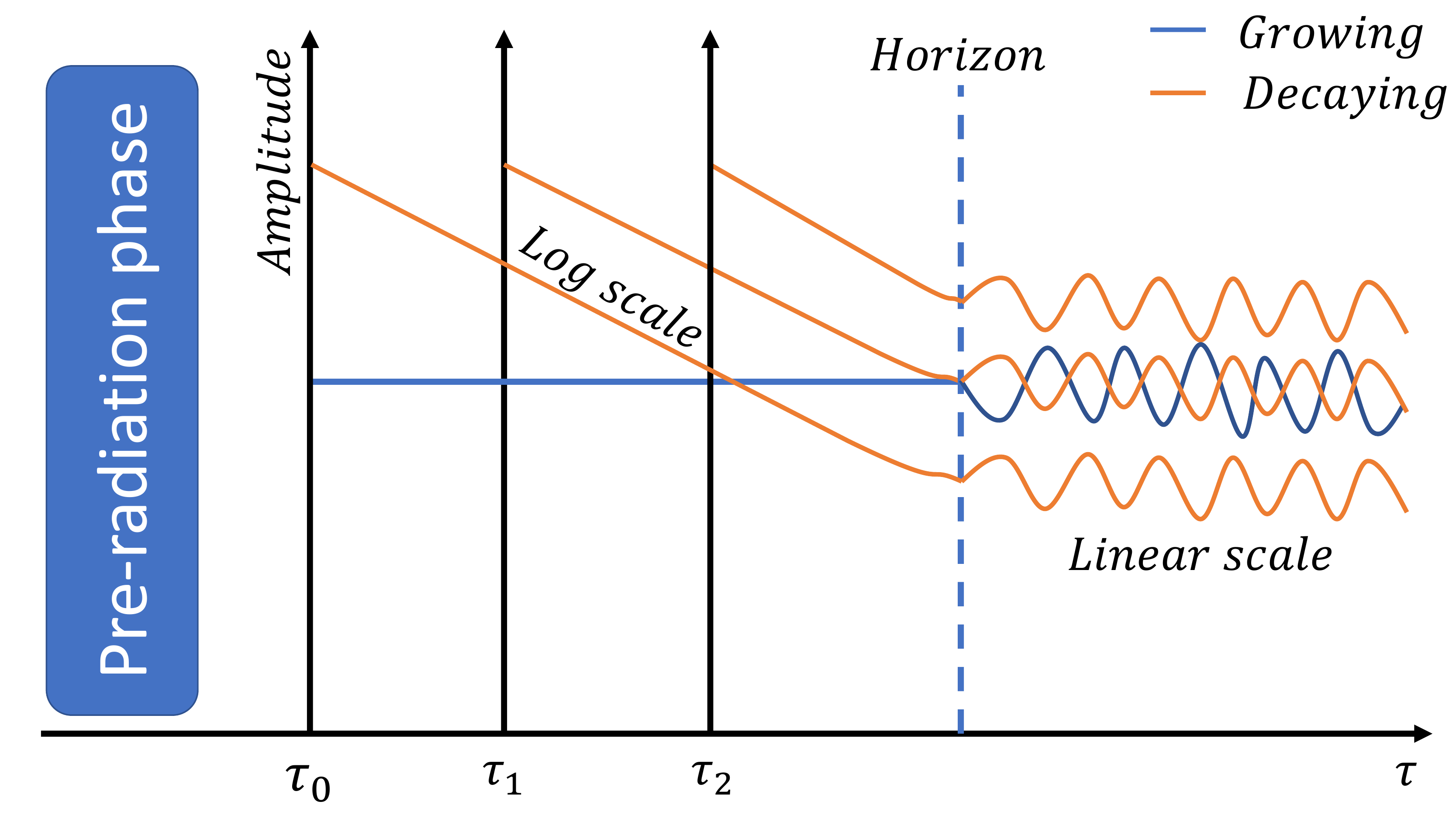}
\caption{Schematic diagram showing that in general both the decaying and growing modes should be sourced by whatever pre-radiation dominated era sets the initial conditions of the universe.
The amplitude of the decaying mode is time-dependent and therefore the amplitude of the decaying mode at recombination is very sensitive to the initial time the amplitudes are set. The amplitude is in log(linear) scale for the super(sub) horizon modes. While the numerical value of the amplitude appears to diverge super-horizon, it does not lead to divergent observable constraints.}
\label{Schematic}
\end{figure}

The decaying mode is qualitatively different to the growing mode as its amplitude is time dependent even on super-horizon scales as shown in Fig.~\ref{Schematic}.
Furthermore, since we are not directly able to measure super-horizon modes it may also be sensible to define these modes by their sub-horizon behaviour. 
On sub-horizon scales, both of these modes are described by oscillatory functions. 
In a pure radiation Universe, the decaying solution is a sine wave and the growing solution is a cosine wave. 
We will use the names {\it sine(cosine)} modes or {\it decaying(growing)} modes interchangeably throughout this paper.
While it is difficult to source decaying modes from inflation, there are scenarios in which they might be generated. 
Specifically, there have been many studies of bouncing and cyclic universes in which decaying modes can be sourced. 
In particular growing modes in a pre-bounce contracting phase can become decaying modes in the post-bounce expanding phase \cite{ Bozza:2005xs, Battefeld:2005cj, Chu:2006wc, Brandenberger:2007by, Alexander:2007zm}.
There is currently no consensus on how the modes are matched across a bounce as this involves understanding the quantum behaviour of the fields causing the bounce in the large curvature regime. 
There have been some recent attempts at computing the propagation of perturbations across a bounce both classically and quantum mechanically in \cite{Gielen:2015uaa, Gielen:2016fdb} which suggest decaying modes could be present. 
More recent studies of the perturbations have gone beyond the leading order expansions and have shown that the decaying modes will also be sourced at second order in perturbation theory (for example from the neutrino velocity mode as it sources anisotropic stress) even if at leading order one only keeps growing modes \cite{Carrilho:2018mqy}. 
Instead of studying a particular scenario in detail we instead use the studies above as motivation to study decaying modes in general.

There has only been one study \cite{Amendola:2004rt} which has attempted to analyse the effect of decaying modes and our aim is to further elaborate and build on this analysis.
In this study we quantify how large the amplitude of these decaying modes can be irrespective of how they are sourced.
We do this by finding the Fisher information in each bin of $k$ in the decaying mode power spectrum, similar to what is done in studies that attempt to reconstruct the power spectrum for the growing mode \cite{Gauthier-Bucher, Bridle:2003sa, Hazra:2017joc, Kogo:2003yb, Nicholson:2009zj}. 
This gives a direct handle on the fraction of decaying modes present on all scales in the universe at the time of recombination.
We will show the constraints on the decaying mode power spectrum that come from using both the temperature and polarization angular power spectrum of the CMB.

The paper is organised as follows. 
In section \ref{theory_decay} we present an intuitive explanation for the growing and decaying modes in a pure radiation universe. 
We then extend this analysis to the describe the initial conditions in general in both the Synchronous and Newtonian gauge to analyse the gauge dependence of the gravitational potentials and confirm the time dependent behaviour of decaying modes on super-horizon modes.
With the time dependence established we provide a normalisation procedure of decaying modes on subhorizon modes.
In section \ref{Analysis} we describe our formalism to constrain the power in the decaying modes using a Fisher matrix formalism and present the results.
We conclude and address possible future directions in \ref{conclusion}

\section{Theory of the decaying mode}\label{theory_decay}

\subsection{Review of radiation domination}

The equations that govern the evolution of the perturbations in standard cosmology are the perturbed Einstein equations. 
In homogenous and isotropic models of the universe, the solution to the Einstein equations is given by the Friedmann-Robertson-Lemaitre-Walker (FRLW) metric.
In the Newtonian (N) gauge, the perturbed FRLW metric for scalars is parametrised by
\begin{equation}
	ds^2_{\rm N} = a(\tau)^2\left( - d \tau_{\rm N}^2(1 + 2 \Phi) + dx^i_{\rm N} dx^j_{\rm N} \gamma_{ij}(1 - 2\Phi) \right). \label{Newtonian}
\end{equation}  
Here $a(\tau)$ is the conformal scale factor and $\gamma_{ij}$ is the flat three dimensional metric on spatial hyper-surfaces. 
This parametrisation of the metric is particularly useful to analyse the physical behaviour of perturbations as it is directly related to the gauge invariant Bardeen potentials, $\Psi = \Psi_{\rm B}, \Phi = - \Phi_{\rm B}$ \cite{Bardeen:1980kt}. 
The equation of motion for the gravitational perturbations in the presence of a pure radiation fluid in the Newtonian gauge, in the absence anisotropic stress, is given by \cite{MUKHANOV1992203}
\begin{equation}
	\Phi'' + 3\mathcal{H} (1 + c_s^2) \Phi' - c^2_s \nabla^2 \Phi + (2 \mathcal{H}' + (1 + 3c_s^2) \mathcal{H}^2) \Phi = 4 \pi G a^2 \tau \delta S. \label{phi_EOM}
\end{equation}
Here $\mathcal{H} \equiv a' / a$ is the conformal Hubble parameter and $\delta S$ is a source term (See Eq.~(5.22) in \cite{MUKHANOV1992203} for full definitions). 
The source term is generated by isocurvature fluctuations and thus is zero for a pure adiabatic solution.
If we restrict ourselves to the radiation dominated era of the universe and without isocurvature, Eq.~\eqref{phi_EOM} simplifies in Fourier space to 
\begin{equation}
	\Phi_k'' + \frac{4 \Phi'_k}{\tau} + \frac{k^2 \Phi_k}{3} = 0,
\end{equation}
which has a simple solution
\begin{equation}
	\Phi_k = A_k \frac{j_1(x)}{x} + B_k \frac{n_1(x)}{x}.
\end{equation}
The amplitudes $A_k$ and $B_k$ are set by the initial conditions for the differential equation, which are the initial conditions for our universe. 
The $k$ index shows that the amplitude can be different for different $k$'s.
Here we have defined $x \equiv \frac{k \tau}{\sqrt{3}}$. The $j_1(x)$ and $n_1(x)$ are the Bessel and Neumann functions of order 1 respectively. 
The term with the Bessel (Neumann) function is the {\it growing (decaying)} which have a cosinal and sinusoidal oscillation respectively.  
It is illuminating to look at the asymptotic limit of these modes. 
At early times on super-horizon scales, i.e. $x \ll 1$, the potential becomes 
\begin{equation}
	\Phi_k(x\ll1) = \frac{A_k}{3} + \frac{B_k}{x^3} \label{decay_1}.
\end{equation}
Here we see that the decaying mode diverges as $x \rightarrow 0$. 
Furthermore, in most models of inflation the decaying mode will be suppressed by $\mathcal{O}(e^{3N})$, where $N$ is the number of e-folds, as the curvature perturbations in inflation will have their amplitudes set at a much earlier time. 
These are the main reasons behind most cosmological analysis assuming $B_k =0$. 
We also see that the growing mode is a constant on super-horizon scales. 
The usual procedure is to match the primordial curvature perturbation $\mathcal{R}_k$ to the amplitude of $A_k$, i.e $\mathcal{R}_k(\tau = 0 ) = - \frac{3}{2} \phi_k(\tau=0)$. 
Now lets analyse the large $x$ limit (sub-horizon limit)
\begin{equation}
	\Phi_k(x) = - \left( A_k \frac{\sin{x}}{x^2} + B_k \frac{\cos{x}}{x^2}\right). \label{pure_radiation}
\end{equation}
Here we see that both modes simply oscillate at late times on sub-horizon scales. 
Thus if there was any remaining non-negligible amount of decaying mode amplitude on sub-horizon scales, {\it it would not decay away}. 
It is therefore sensible to ask how large the amplitude of such a decaying mode has to be to lead to observable effects (or similarly, constrained by the data). 
That is the main question we set out to answer in this paper. 

\subsection{CMB anisotropies}

The angular power spectrum of the CMB anisotropies is given by \cite{durrer_cmb}
\begin{equation}
	C^{XY}_{\ell} = \int^\infty_{0} d \ln k \ P^{XY}(k) |\Delta^X_{\ell}(k) \Delta^Y_{\ell}(k)| 
\end{equation}
Here $P(k)$ is the primordial power spectrum of curvature perturbations.  
$X,Y \in \{T, E\}$ where $T,E$ stand for temperature and polarization respectively. 
$\Delta^X_{\ell}(k)$ is either temperature or polarization transfer function for adiabatic modes. 
In general, the transfer functions are computed using a line of sight approach by separating out the geometric projection effects (that depend on $\ell$) and the physical effects coming from gravitational potentials and Doppler effects \cite{Seljak:1996is}. 
On large scales the source function for temperature anisotropies is given by the gravitational potential, $\frac{\Delta T}{T} \approx \frac{1}{3} \Phi$.
This effect is caused by photons from the CMB having to climb out of a gravitational well and is called the Sachs-Wolfe (SW) effect. 
Thus, on large scales the CMB power spectrum should directly see a change in the gravitational potential, such as the change due to decaying modes in Eq.~\eqref{decay_1}.

\begin{figure*}
\begin{center}
\includegraphics[scale=0.45]{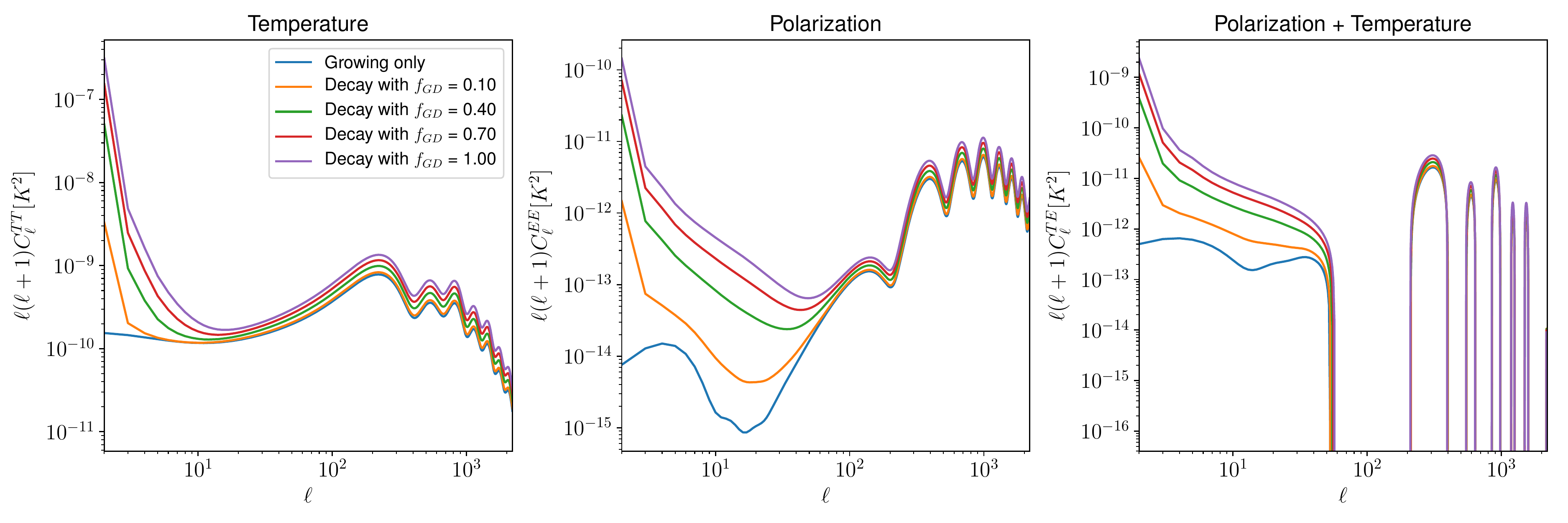}
\caption{Angular power spectrum of the CMB temperature and polarization anisotropies.}
\label{decay_cls}
\end{center}
\end{figure*}

We can check this explicitly by implementing the initial conditions for the decaying mode into the Boltzmann-solver CLASS \cite{CLASSII} and in the synchronous (S) gauge these are parametrised by
\begin{equation}
	ds^2_{\rm S} = a^2(\tau) \left( - d\tau_{\rm S}^2 + dx_{\rm S}^i dx_{\rm S}^j \left(\gamma^{\rm S}_{ij} + h_{ij} \right) \right). \label{Synchronous}
\end{equation}
We will focus on scalar perturbations in this paper and it is canonical to separate $h_{ij}$ into two scalars: its trace $h$ and traceless $6 \eta$ parts.
The initial conditions in this gauge are given by \cite{Amendola:2004rt}
\begin{widetext}
\begin{eqnarray}
h(x, \phi) &=& x^2 + f_{\rm GD}x^\frac{3}{2} \sin{\xi}, 	\nonumber	\\
\eta(x, \phi) &=&  2 - \frac{5 + 4R_\nu}{6(15+4R_\nu)} x^2  +  \frac{f_{\rm GD}}{x^\frac{1}{2}} \left[ \frac{11 - \frac{16R_\nu}{5}}{8} \sin{\xi} + \frac{5 \gamma}{8} \cos{\xi} \right],	\nonumber	\\
\delta_\nu(x, \phi)  &=& - \frac{2x^2}{3} + f_{\rm GD}x^\frac{3}{2} \left[ \left( \frac{1}{4R_\nu} - \frac{2}{5} \right) \sin{\xi} - \frac{\gamma}{4R_\nu} \cos{\xi} \right],	\nonumber	\\
\Theta_\nu(x, \phi)  &=& - \frac{23 + 4R_\nu}{18(15+4R_\nu)} kx^3 + \frac{f_{\rm GD}}{16R_\nu}kx^\frac{1}{2} \left[ \left( -3 - \frac{72}{5} R_\nu \right) \sin{\xi} + \gamma \left(3 - \frac{8R_\nu}{5} \right) \cos{\xi} \right],	\nonumber	\\
\Theta_r(x, \phi)  &=& \Theta_b = - \frac{kx^3}{18} + \frac{f_{\rm GD}kx^\frac{5}{2}}{3(25+\gamma^2)} \left(\gamma \cos{\xi} - 5 \sin{\xi} \right), \nonumber	\\
\sigma_\nu(x, \phi)  &=& \frac{4}{3(15+4R_\nu)}x^2 + \frac{f_{\rm GD}}{x^\frac{1}{2}} \left[ \frac{\gamma}{2} \cos{\xi} + \frac{11 - 16R_\nu/5}{10} \sin{\xi} \right],	\nonumber	\\
\delta_r(x, \phi)  &=& - \frac{2}{3} x^2 - \frac{2f_{\rm GD}}{3} x^\frac{3}{2} \sin{\xi},	\nonumber	\\
\delta_c(x,\phi)  &=& \delta_b = - \frac{x^2}{2} - \frac{f_{\rm GD}x^\frac{3}{2}}{2} \sin{\xi},
\label{decay_ic}
\end{eqnarray}
\end{widetext}
with the following definitions
\begin{eqnarray}
	\xi & \equiv & \frac{\gamma}{2} \log{x} + \phi; \,\,\,\gamma  \equiv  \sqrt{\frac{32}{5} R_\nu -1},	\nonumber	\\
	x & \equiv & k\tau;\,\,\,\,\,\,\,\,\,\,\,\,\,\,\,\,\,\,\,\,\,\,\,R_\nu  \equiv  \frac{\rho_\nu}{\rho_\nu + \rho_\gamma}.
\end{eqnarray} 
The amplitude $f_{\rm GD}$ is the ratio of the decaying mode to the growing mode.
We have defined the densities $\delta_i$, velocities $\Theta_i$ for each of the species $i \in \{$radiation (r), CDM (c), Baryons (b), neutrinos $(\nu)\}$. 
$\sigma_\nu$ is the quadrupole moment of the neutrino phase space density and $R_\nu$ is the relative energy density fraction of neutrinos.
The physical reason for the neutrinos having a quadrupole is that they will have anisotropic stress after they decouple. 
However this is also the case for the growing adiabatic mode \cite{Bucher:1999re, Ma:1995ey}, which can be obtained by setting $f_{\rm GD}$ equal to zero in the Eq.~\eqref{decay_ic}. 
We also note that the decaying mode has two independent variables $f_{\rm GD}$ and $\phi$. 
This is because for decaying modes there is an additional equation of motion for the neutrino distribution. 
This can easily be seen if one considers a pure radiation fluid coupled to neutrinos, as was pointed out in \cite{Amendola:2004rt}. 
It is known that the growing mode remains constant on super-horizon scales. 
However this is not the case for decaying modes. 
We have already seen this for a pure radiation universe in Eq \eqref{pure_radiation}.
The initial conditions in the synchronus gauge do not make the time dependence(or independence) apparent as it appears both growing and decaying modes are time dependent. 
However the metric potentials $\eta$ and $h$ are not gauge invariant quantities. 
It is, therefore, better to analyse the time dependence in the Newtonian gauge as the metric potentials are directly related to the gauge invariant Bardeen potentials. 
We can switch to Newtonian gauge by either solving the Boltzmann equations in the Newtonian gauge or, as we are only interested in the behaviour of the gravitational perturbations, we can relate the two metrics via, 
\begin{equation}
	g_{\mu \nu(\rm N)} = g_{\alpha \beta(\rm S)} \frac{\partial x^\alpha_{(\rm S)}}{\partial x^\mu_{(\rm N)}} \frac{\partial x^\beta_{(\rm S)}}{\partial x^\nu_{(\rm N)}},
\end{equation}
where the variables with $(\rm N) / (\rm S)$ are in the Newtonian / Synchronous gauge which are defined in Eq.~\eqref{Newtonian}, \eqref{Synchronous} respectively. 
The relations between the metric potentials can then be calculated to be
\begin{eqnarray}
	\Psi(x, \phi) & = & \frac{1}{2k^2} \left[ \ddot{h}(x, \phi) + 6 \ddot{\eta}(x, \phi) + \frac{\dot{a}(\tau)}{a(\tau)} \left( \dot{h}(x, \phi) + 6 \dot{\eta}(x, \phi) \right) \right], \nonumber \\
	\Phi(x, \phi) & = & \eta(x, \phi) - \frac{1}{2k^2} \frac{\dot{a}(\tau)}{a(\tau)} \left[ \dot{h}(x, \phi) + 6 \dot{\eta}(x, \phi) \right].
\end{eqnarray}
Using these to evaluate the Newtonian potentials we get
\begin{widetext}
\begin{eqnarray}
	\Psi(x, \phi) &=& \frac{20}{15 + 4R_\nu} + \frac{f_{GD}}{8 x^{\frac{1}{2}}} \left( 6 \gamma \cos \xi - (9-\gamma^2) \sin \xi \right)  + \mathcal{O}(x^{-\frac{5}{2}}) \nonumber \\
	\Phi(x, \phi) &=& \frac{4(5+2R_{\nu})}{15 + 4 R_\nu} + \frac{f_{GD}}{40 x^\frac{1}{2}} \left[(15 \gamma \cos \xi + (25 - 16R_\nu) \sin \xi  \right) + \mathcal{O}(x^{-\frac{5}{2}}) 
    \label{Newtonian_pot}
\end{eqnarray}
\end{widetext}
We see that for the growing mode, i.e when $f_{\rm GD} = 0$, the metric potentials are constant. 
Whereas for the decaying mode, the potentials are clearly time dependent. 
Thus, we need to specify the time at which the decaying modes start evolving as the constraints we get on the amplitude will be depend on this time, as is shown in Fig.~\ref{Schematic}. To get an idea of what the power spectrum of the decaying mode looks like we have implemented these initial conditions in the CLASS Boltzmann code and the resulting power spectra for the temperature, polarization and their cross spectrum are shown in Fig.~\ref{decay_cls}.
In Fig.~\ref{decay_cls} we have assumed a power spectrum of the decaying mode to be analogous to the growing mode and set the spectral index $n_s^{\rm D} = n_s^{\rm G} = 0.96$ while the amplitude is defined by the scalar amplitude $A_s^{\rm G}$ and the fraction of decaying mode amplitude $A_s^{\rm D} = f_{\rm GD} A_s^{\rm G}$. This amplitude will be time dependent as explained before, so we set it at the minimum time used in the source function integral\footnote{A summary of schemes used for setting the initial conditions is given in Fig.~10 of \cite{CLASSII}.}. Later on when we are computing the errors on the amplitudes we normalise the modes such that the decaying modes have the same amplitude as the growing modes on subhorizon scale.
We set $\phi = 0$ and the rest of the cosmological parameters are set to the fiducial values given in table \ref{fid_params}.

\begin{figure}
	\centering 
	\includegraphics[scale = 0.35]{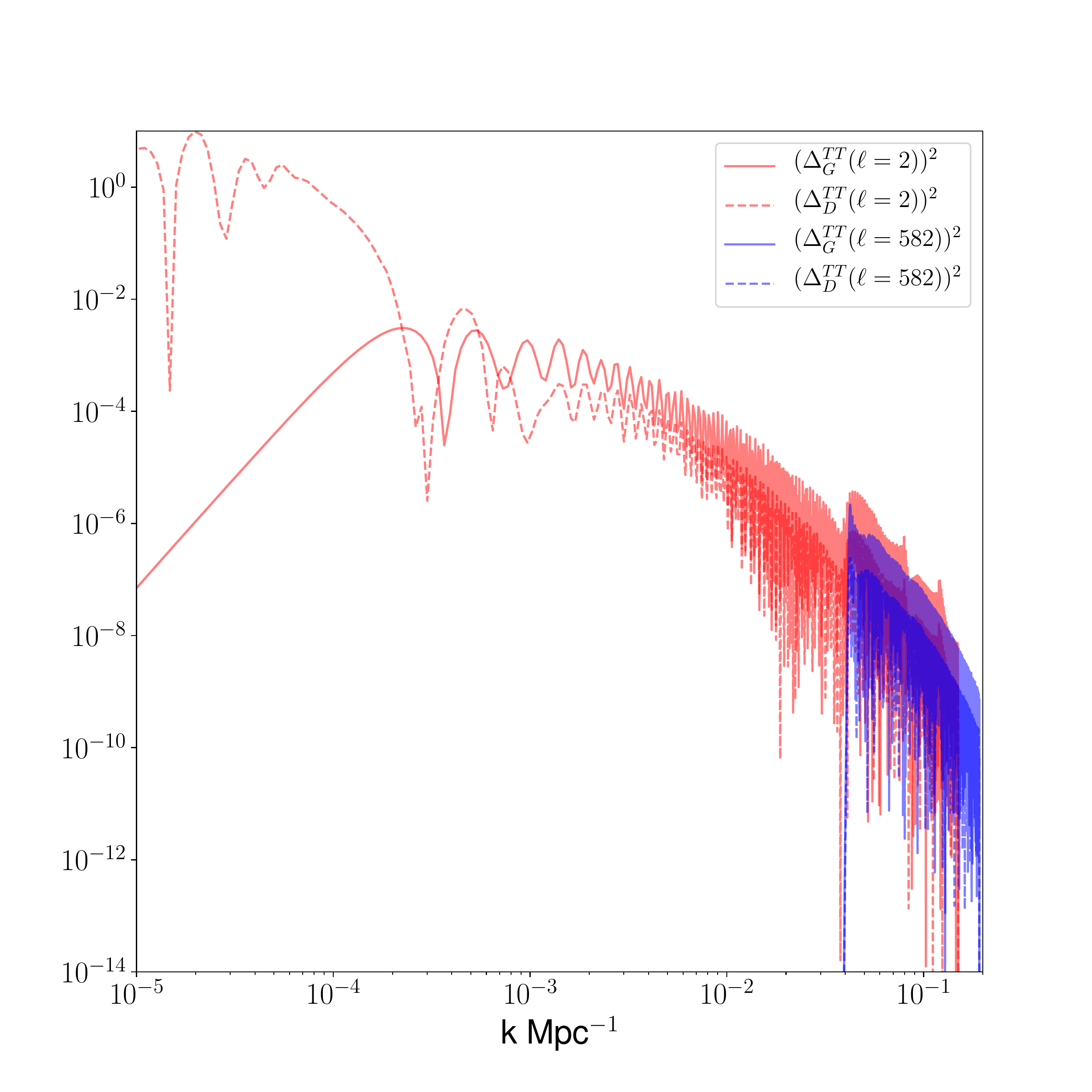}
	\caption{Transfer functions for growing and decaying modes.}
	\label{transfer_plots}
\end{figure}

\begin{center}
\begin{table}[h!]
\begin{tabular}{ p{1.5cm}  p{2cm}}
\hline
\hline
$A_s$ & 2.3 $\times 10^{-9}$  \\
$h$ & 0.6711  \\
$\Omega_bh^2$ & 0.022068  \\
$\Omega_{cdm}h^2$ & 0.12029 \\
$k_{*}$ & 0.05 $Mpc^{-1}$ \\
$n_s $ & 0.9619 \\
$N_{eff}$ & 3.046 \\
\hline
$\ell_{max}$ & 2500 \\
$f_{sky}$ & 1 \\
\hline
\hline
\end{tabular}
\caption{Fiducial cosmological parameters and systematic parameters}\label{fid_params}
\end{table}
\end{center}
There is a clear divergence on large scales which comes from the divergence of the gravitational potential on super-horizon scales. 
The gravitational potential enters the $C_{\ell}$'s through the transfer function's $\Delta_{\ell}(k)$.
These are (numerically) computed using a line of sight integral \cite{Seljak:1996is} over the source function (which contains the Sachs Wolfe, Doppler and Integrated Sachs Wolfe effect terms) convolved with a projection function which is a Bessel function. 
\begin{equation}
	\Delta_{\ell}(k) \equiv \int^{\tau_0}_{\tau_i}  \ d \tau \ S_T(\tau, k) j_{\ell}(k(\tau_0 - \tau))
\end{equation} 
Here $\tau_0$ is the time at recombination and $\tau_i$ is the time at which the initial conditions are sourced. We show the transfer functions for $\ell =2, 582$ in Fig.~\ref{transfer_plots}. 

The low $\ell$'s show the divergent behaviour for the decaying mode, whereas at $\ell = 582$ we see that both modes are similar with the decaying mode having a lower amplitude. 
This means a non-negligible amplitude of the adiabatic perturbations could be in decaying modes if they are generated at late times or on large scales. 
Furthermore a primordial power spectrum with a large spectral index could also allow for a non-negligible contribution of the decaying mode amplitude to the overall adiabatic perturbations. 

Instead of focusing on setting the amplitude at early times, we use a renormalising procedure to set the amplitude of the decaying modes.
There are two reasons to use this normalisation procedure. 
First, it provides a unique way to set the initial conditions as the decaying modes are time dependent and the time dependence is \emph{different} in different gauges. 
For example the time dependence of decaying mode metric potentials in the Synchronous gauge in Eq.~\eqref{decay_ic} is clearly different from the metric potentials in the Newtonian gauge in Eq.~\eqref{Newtonian_pot}.
Second, since both the growing and decaying solutions are described by regular (non-diverging) functions on sub-horizon scales we can set the amplitudes of the growing mode equal to that of the decaying mode deep inside the horizon.
This makes it easier to see the effect of decaying modes that are set at late times as they would naturally be normalised on sub-horizon scales. 

\begin{figure}
	\centering 
	\includegraphics[scale = 0.3]{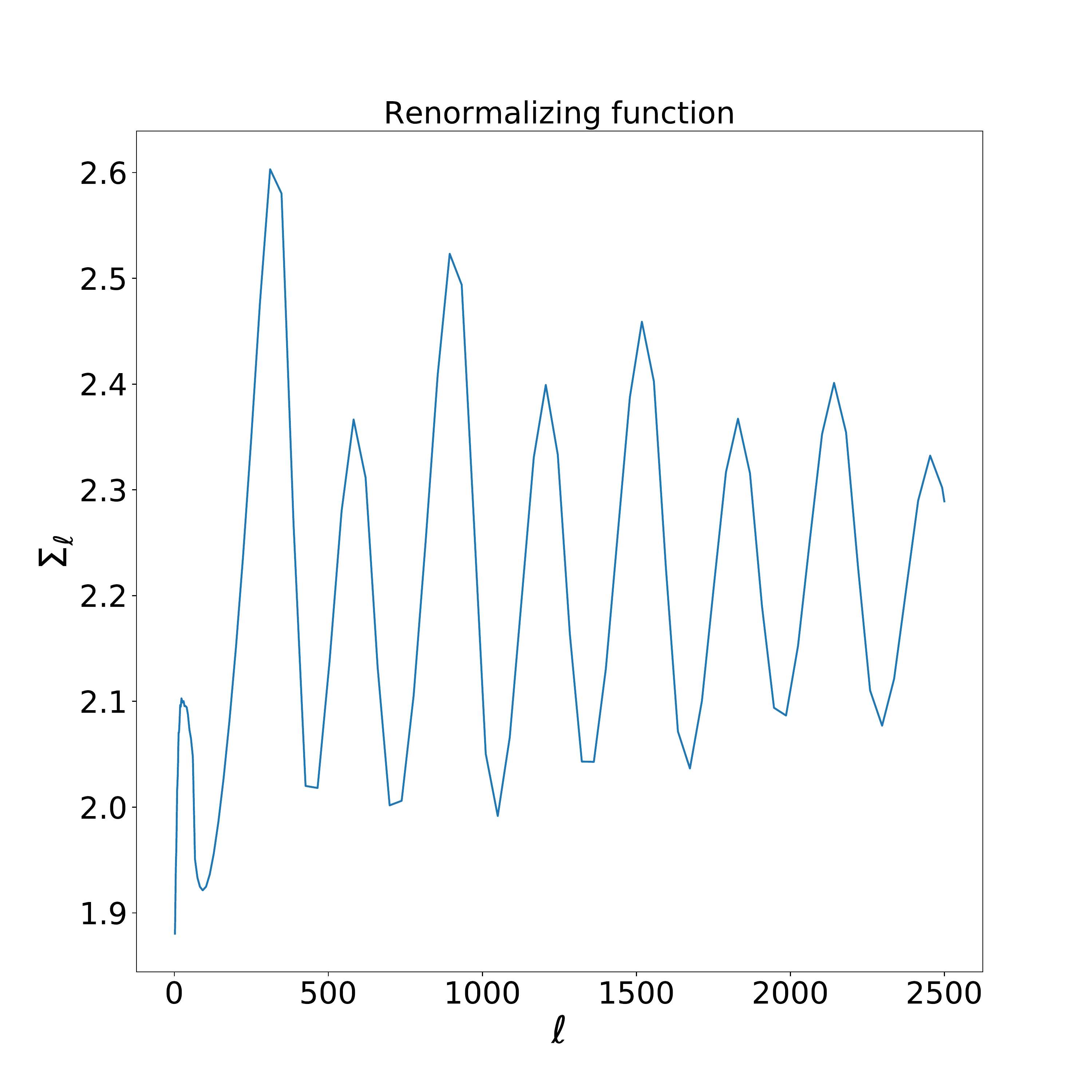}
	\caption{Renormalisation function defined in Eq (\ref{renorm_func})}
	\label{renorm}
\end{figure}

The normalisation of the two modes is done in terms of the transfer functions in $k$ space as opposed to the transfer functions in $\ell$ space as we wish to isolate the physical effects of the gravitational potentials (which show the behaviour of the growing and decaying modes) from the projection effects.
We equate the amplitudes of the decaying and growing modes on all scales below the fiducial horizon scale $k_{\rm horizon} = 3 \times 10^{-3} \mbox{ Mpc}^{-1}$. 
In practice it is not easy to do this since the transfer functions are highly oscillating functions.
Our approach is to integrate the transfer function for each $\ell$ for all $k$'s that are inside the horizon for both the growing mode and decaying mode. 
The ratio of these integrals will tell us the normalisation for the decaying mode transfer function for a given $\ell$ that will ensure the decaying mode will have the same amplitude as the growing mode on sub-horizon scales. 
This would correspond to the case where the universe starts at $\tau_1$ in Fig.~\ref{Schematic}.
Thus the renormalised decaying mode transfer function can be written as 
\begin{eqnarray}
	\hat{\Delta}^D_{\ell}(k) & = & \Delta^D_{\ell}(k) \Sigma_{\ell}, \nonumber \\
	\Sigma_l & \equiv &  \frac{ \int_{k_{\rm horizon}}^{k_{\rm max}} dk \ \Delta^G_{\ell}(k)}{\int_{k_{\rm horizon}}^{k_{\rm max}} dk \ \Delta^D_{\ell}(k)}. \label{renorm_func}
\end{eqnarray}
We know \emph{a-priori} that this is the most conservative one can be as for decaying modes to have the same amplitude as growing modes on sub-horizon scales they must have a very large amplitude on super-horizon scales (at least for modes that entered that horizon at early times) and thus they will be highly constrained. 
Any early universe model that is responsible for generating the initial conditions can be renormalised in this way, thus allowing a direct comparison of the amplitudes of a model to our results by simply applying the renormalization function in Fig.~\ref{renorm}. 

\section{Analysis}\label{Analysis}

\begin{figure}
	\centering 
	\includegraphics[scale = 0.25]{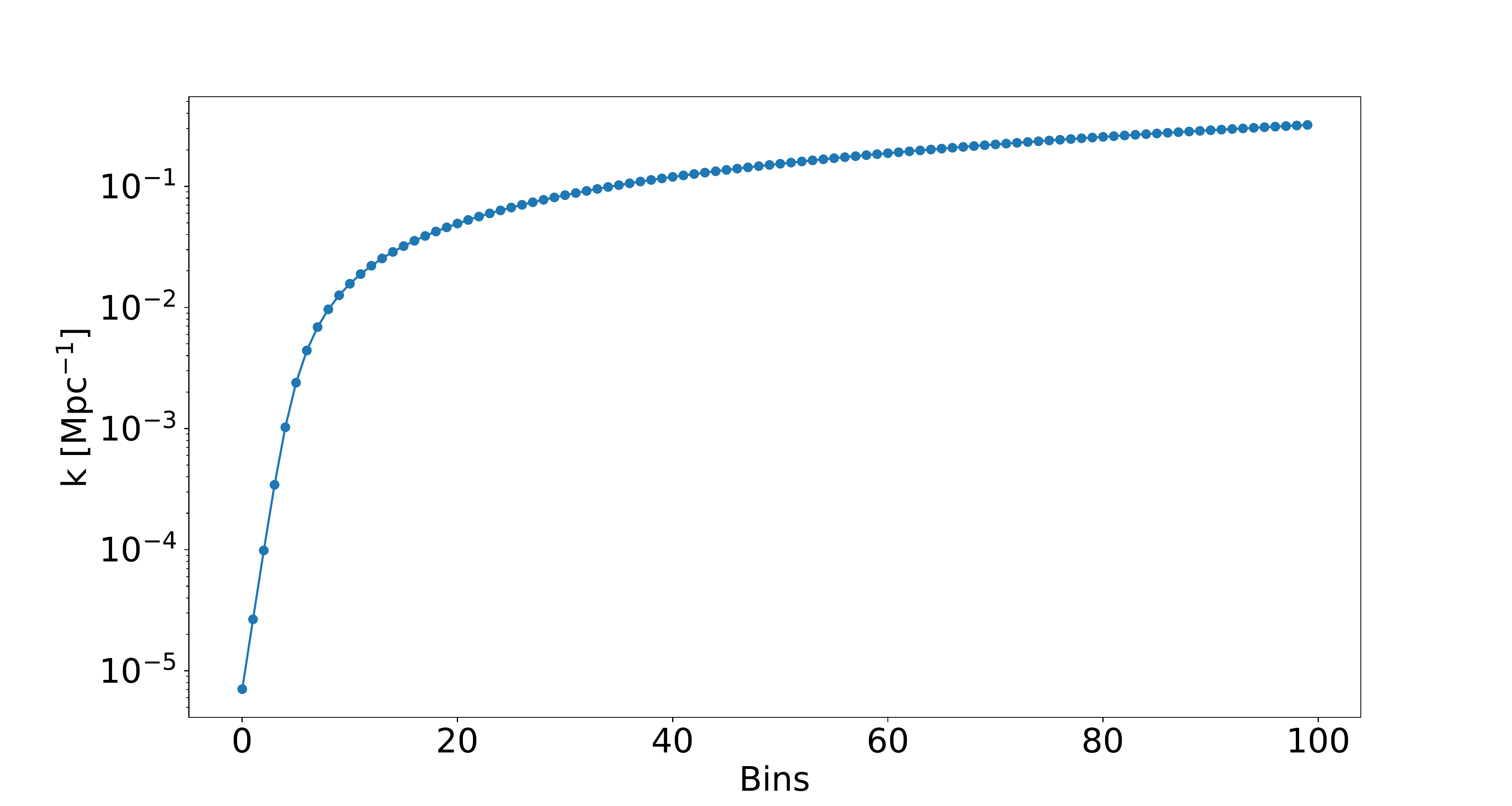}
	\caption{$k$ values at which we add power to the primordial power spectrum}
	\label{kbins}
\end{figure}

There are a variety of ways to model the primordial power spectrum. The most popular one, and the one which is normally constrained with data, is a power law with an amplitude and spectral index. There are a variety of ways to look for deviations from this. Here we take an unparameterised approach to constraining the decaying mode to keep our findings as general as possible.   
For that purpose, we model the power spectrum as a set of bins in $k$ with an independent amplitude and constrain the amplitude in each of those bins. 
The power spectrum is then given by 
\begin{eqnarray}
P(k,k_0,\epsilon) = \begin{cases}
P(k)^{(G)} + \epsilon^{(G) \mbox{ or } (D)}_{k_0} & \text{if $k_0\equiv k $} \\
P(k)^{(G)} & \text{otherwise}
\end{cases} \nonumber \\
\end{eqnarray}
where $P(k)^{(G)} = A^{(G)}_s \left( \frac{k}{k_*} \right)^{n_s^{(G)} - 1}$. 
We choose 100 values for $k_0$ from an infrared cutoff of $3 \times 10^{-5} \ \mbox{Mpc}^{-1}$ to $3 \times10^{-1} \ \mbox{Mpc}^{-1}$ with the precise values for each bin shown in Fig.~\ref{kbins}. 

To account for the information on smaller scales we would also need to account for CMB lensing due to large scale structure which we know can change the temperature power spectrum by $\mathcal{O}(20 \%)$ on scales below $\ell \sim 3000$, thus we do not look at smaller $\ell$'s.
This parametrisation allows us to look for features in the primordial power spectrum that can arise by either the growing mode or the decaying mode. 
In the case where the feature is due to the decaying mode, i.e $\epsilon^{(D)}$ is added to the power spectrum, we also use the decaying mode transfer functions to evaluate the $C_{\ell}$'s.
Since the $C_{\ell}$'s are a linear function of the power spectrum, the total $C_{\ell}$'s will just be the sum of the growing mode fiducial power spectrum $C_{\ell}$'s and a response due to the decaying mode being added. 
We will also consider the effect adding polarization information has on the constraints.
Since the transfer function for the decaying mode is different for polarization and temperature, the same primordial power spectrum may not be able to account for the change in temperature and polarization. 
A similar analysis has been done for parametrised isocurvature modes \cite{Bucher:2000cd} and it was shown that adding polarization significantly increases the constraining power of the CMB for the amplitude of isocurvature modes. 
In principle one could apply this un-parametrised approach to primordial isocurvature perturbations as well and we will leave this to future works.

\begin{figure*}
\begin{center}
\includegraphics[scale=0.5]{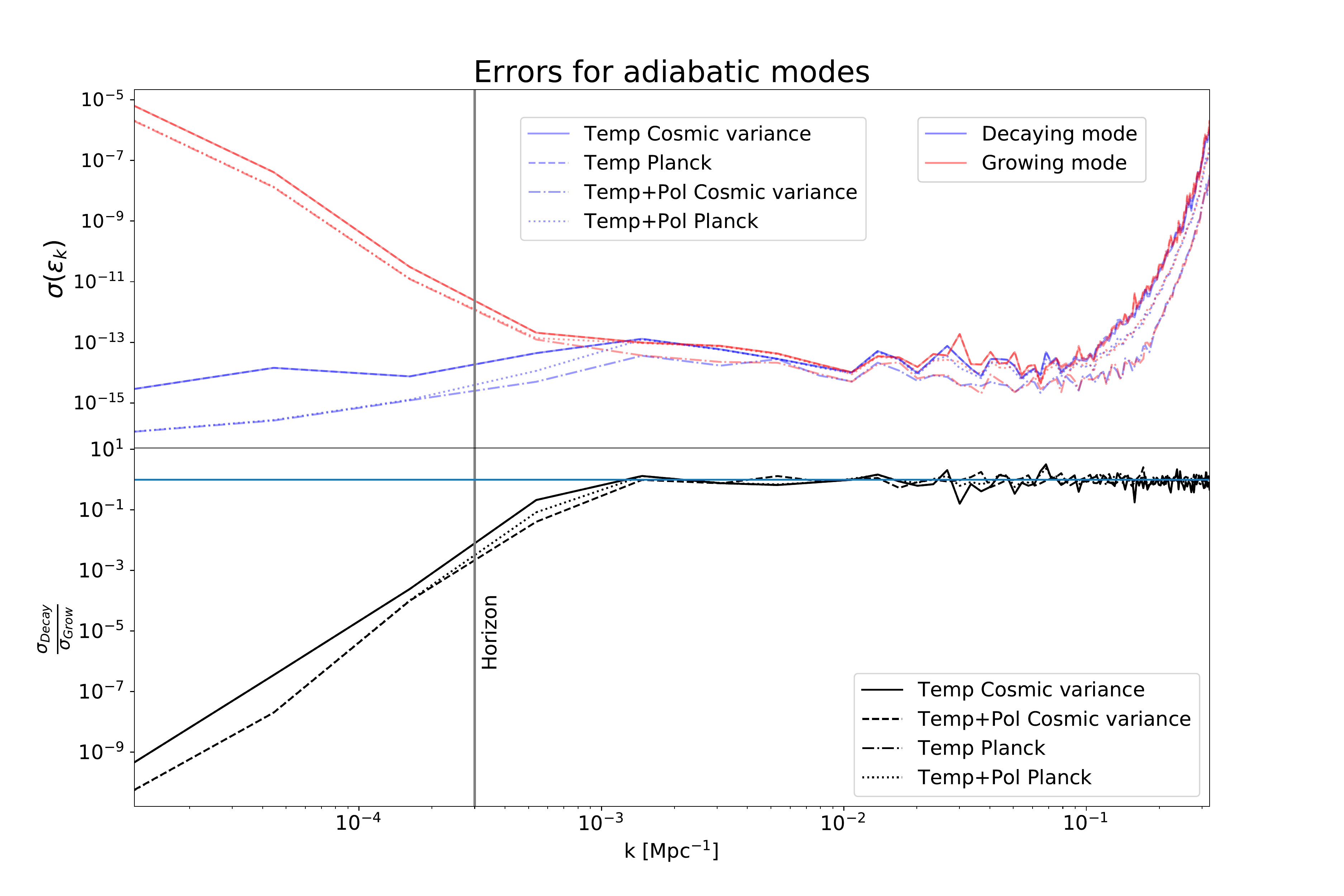}
\caption{This plot shows the errors for the decaying and growing modes in each of the 100 $k$ bins. The analysis is done for four specifications:  temperature anisotropies in a Planck like experiment and a cosmic variance limited experiment and the same analysis for temperature and polarisation data. The top plot shows the errors and the bottom plot shows the ratio of the errors of the decaying to growing modes. The vertical line is drawn at roughly the size of the horizon as inferred from the maximum scale observable by an observer at recombination. Since $\ell = 2$ is the largest mode observable in the CMB we compute the corresponding $k$ using $\ell = k \chi$. This expression is true in a flat sky, where $\chi$ is the comoving distance to recombination $\sim 10$ Gpc/h, furthermore $\ell =2$ corresponds to a mode wave with two wavelengths in a unit circle giving, thus giving a further factor of $\pi/2$ to the wavenumber giving $k\sim 3\times10^{-4} \mbox{ Mpc}^{-1}$. The horizontal line on the bottom plot is at one and we note that the ratio of the errors asymptotes to 1. This is just a manifestation of the fact that we have normalised the amplitudes (but not the phase) of both modes to be equal on sub-horizon scales.}
\label{fish_sigma}
\end{center}
\end{figure*}

To answer these questions we use the Fisher information as a metric to quantify the information in the decaying modes. 
The expression for the Fisher matrix for a Gaussian likelihood with a parameter independent covariance matrix can be written as 
\begin{equation}
	F_{\alpha \beta} = \sum_{l=2}^{l_{max}} \frac{f_{sky}(2l+1)}{2} \text{Tr} \left( \mathbb{C}_l^{-1} \partial_\alpha \mathbb{C}_l \mathbb{C}_l^{-1} \partial_\beta \mathbb{C}_l \right).
\end{equation}
The matrix $\mathbb{C}$ depends on the observables being used. 
When the temperature and polarization of the CMB are being used the matrix becomes
\begin{equation}
	\mathbb{C}_l \equiv \begin{pmatrix} C_l^{TT} + N_l^{TT} & C_l^{TE} \\ C_l^{ET} & C_l^{EE} + N^{EE}_l \end{pmatrix}.
\end{equation}
The fiducial $\mathbb{C}_l$ is assumed to be that of the growing mode only as we know it fits the data with the fiducial cosmology. 
The derivatives of the $\mathbb{C}_l$ matrix will have either the growing or decaying transfer functions, depending on which mode is being constrained.
Where $N_l^{TT}, N_l^{EE}$ represent the noise covariance for temperature, polarization respectively. 
We also assume the polarization and temperature noise are uncorrelated thus the covariance between them is zero.
We model the noise for the CMB polarization and temperature as Gaussian random noise per frequency channel as given in the Planck blue book \cite{Planck:2006aa}
\begin{eqnarray}
	N^{TT(EE)}_l & = &  \left( (\sigma^2_{T(E)}B^2_l)_{100} + (\sigma^2_{T(E)}B^2_l)_{143}  \right. \nonumber \\
	& + & \left. (\sigma^2_{T(E)}B^2_l)_{217} + (\sigma^2_{T(E)}B^2_l)_{353} \right)^{-1}
\end{eqnarray} 
where $\sigma_{T(E)}$ represent the variance for temperature (polarization) and $100,143,217,353$ are the Planck frequency channels in $GHz$. 
The window function is given by $B^2_l = \mbox{exp} \left({-\frac{l(l+1) \theta^2_{beam}}{8 \ln 2}} \right) $. 
The values of the beam size and variance are given in Tab.~\ref{Planck_noise}.

\begin{center}
\begin{table}[h!]
\begin{tabular}{ |p{1.5cm}p{2cm}p{2 cm}p{2 cm}|}
\hline
Frequency ($GHz$) & $\theta_{beam} (rad)$ &  $\sigma_T$($\mu K$ - rad) & $\sigma_E$($\mu K$ - rad)   \\
\hline
100 & 0.002763 & 0.001984 & 0.003174 \\
143 & 0.002065 & 0.001746 & 0.003333 \\
217 & 0.001454 & 0.003809 & 0.007785 \\
353 & 0.001454 & 0.011665 & 0.023647 \\
\hline
\end{tabular}
\caption{Planck noise}\label{Planck_noise}
\end{table}
\end{center}

If we only use the temperature spectrum from the CMB the expression for the Fisher matrix simplifies to 
\begin{equation}
	F_{\alpha \beta} = \sum_{l=2}^{l_{max}} f_{sky} \frac{2l + 1}{2} \frac{\partial_\alpha C^{TT}_l \partial_\beta C^{TT}_l}{(C_l^{TT} +N_l^{TT})^2}.
\end{equation}
It is worth noting that the derivatives of the $C_l$'s wrt the parameters $\epsilon^{(D)}_{k_0}/\epsilon^{(G)}_{k_0}$ will simply return the transfer function squared of the decaying/growing mode at $k_0$. 
The errors on the parameters $i$, $\sigma_i$ (which in our case will be the amplitudes in each $k$ bin) can be obtained by $\sigma_i = \sqrt{(F^{-1})_{ii}}$. 
We plot these variances in Fig.~\ref{fish_sigma} along with the ratio of the errors of the decaying and growing modes.
We see that most of the information is in the range $ k \sim 10^{-3}- 10^{-1} \mbox{Mpc}^{-1}$ and 
adding the polarization data increases the information content by up to 2 orders of magnitude in this range. 
Similar results for the growing mode have been found in previous studies, see for example \cite{Gauthier-Bucher, Mukherjee:2005dc, Hu:2003vp}. We note that most of the analysis done so far focus on providing detailed precision on growing modes and thus have more $k$ bins in a narrower range of wavenumber. 
Our aim is to probe the errors on a much broader range of $k$'s which has not been done before, yet we still note that in regions of overlapping $k$ space we recover similar results albeit without the same level of resolution. 

We see that on larger scales cosmic variance dominates and most of the information is lost.
The first thing to note about the decaying mode is that the overall difference in the Fisher information from the largest to the smallest scales is much lower than the growing mode.
This is because on large scales we can see from the $C_l$'s there is a large rise in power for the decaying mode transfer functions. 
Therefore even with the large errors due to cosmic variance, the excessive power in decaying modes on large scales can be constrained.  
On subhorizon scales the errors on both modes are approximately the same as we have normalised both modes to have the same amplitudes on subhorizon scales.
The second feature of the decaying mode is that there is a large increase in Fisher information, relative to the growing mode, when polarization information is included. 
This is to be expected because, as was mentioned before, the polarization transfer functions and temperature transfer functions are different. 
The fiducial cosmology we have assumed has been fitted to the temperature and polarization data with growing mode transfer functions, thus even if we allow a lot of freedom in the primordial power spectrum, the $C_l$'s, which are a convolution between the transfer functions and the primordial power spectrum, will struggle to accommodate the decaying mode power spectrum with the temperature and polarization transfer functions at the same time.

Finally it is interesting to note that modes that are smaller than $ \sim 10^{-4}$ Mpc$^{-1}$ will be larger than the universe's horizon today and some modes that are even larger may never enter the horizon of our universe. 
Thus one has to ask the philosophical question of how modes that are beyond our observable universe can be observed, even indirectly. 
The physical mechanism for super-horizon modes effecting sub-horizon observables is either through the gravitational effect of super-horizon modes on small scale structure or, potentially the dominant effect, through the effect of spatial gradients of the density perturbations. 
A discussion of the gradient method to analyse long wavelength perturbations can be found in \cite{Harada:2015yda, Lyth:2004gb}.
Both of these effects have been at the heart of \emph{separate universe approach} of describing super-horizon perturbations in which the local, sub-horizon, modes evolve in a different \emph{universe} with different cosmological parameters such as curvature, Hubble rate etc.  
Such claims have to be backed up with careful analysis of the underlying physics, in particular the curvature of spacelike surfaces, as one has to understand how the equivalence principle, which would suggest large scale modes should not effect the curvature of spacelike surfaces, can allow for such super-horizon modes to effect the sub-horizon modes. 
There have been many attempts to address this issue and a long yet in-exhaustive list is given here \cite{Rigopoulos:2003ak, Lyth:2003im, Martineau:2005aa, Sasaki:2005ju, Tanaka:2006zp, Takamizu:2008ra}. 
Most of these attempts have focused on calculating the back-reaction of the \emph{growing super-horizon modes} through the non-linear evolution of the modes due to Einstein's equation.
It would be interesting to see whether similar calculations can be used to evolve decaying modes and understand the physical origin on their effect on sub-horizon scales.
We do not attempt to address this here and note that our current study will provide a direct way to test whether the methods used to understand super-horizon evolution of modes lead to testable predictions. 

\section{Summary and future outlook}\label{conclusion}

In this paper we have analysed the constraints on the amplitudes of the primordial power spectrum across a broad range of scales for adiabatic initial conditions.
Adiabatic initial conditions have two orthogonal set of modes that can be excited when the universe starts (during radiation domination) or at later times. 
These are the sine (decaying) mode or the cosine (growing) mode.
In general both modes can be excited however most cosmological analysis assume only the cosine mode is excited and thus the constraints on the amplitudes of the primordial power spectrum is directly matched to the amplitude of the cosine mode. 

The sine mode numerically appears to diverge at early times on super-horizon scales. Special care is needed to interpret super-horizon physics, and a mapping onto physical quantities is essential.
Past work attempted to normalize the decaying mode at a super-horizon initial condition, making the allowed amplitudes for the sine mode sensitive to the numerical start time universe.
Instead of taking a parametrised approach, in this analysis we have mapped the amplitude of the primordial power spectrum to the amplitude of both the modes by looking for additional power spectrum features for discrete scales.

We have calculated the Fisher information for both the sine and cosine modes using a fiducial cosmology.
The initial conditions for this cosmology are normalised to be equal for both modes on sub-horizon scales. 
We have computed the Fisher information for these modes for a cosmic variance limited experiment as well as a full sky Planck like experiment with temperature and polarization anisotropies.
Both of the modes are best constrained on scales $k \sim 10^{-3} - 10^{-1}$ Mpc$^{-1}$. 
The sine mode is almost equally well constrained on larger scales, $\sim 10^{-4} \mbox{Mpc}^{-1}$ due to the divergent growth of its amplitude, whereas the cosine mode is less well constrained on these scales as they are cosmic variance limited.
The angular power spectrum for the anisotropies of the CMB are a convolution between the primordial power spectrum and the transfer function. Therefore allowing the primordial power spectrum to be a freely varying function may allow the decaying mode to fit the observed temperature anisotropies, it is unable to fit the polarization anisotropies at the same time as they have different transfer functions. It is worth emphasising that this argument only holds when we keep the cosmological parameters fixed. If we let the cosmological parameter vary \emph{at the same time} as varying the primordial parameters one may be able to find new points in parameter space that fit the observed data that allow for non-negligable amounts of power in the sine mode.

This approach of constraining the initial conditions of the universe can be very useful in understanding the early universe models that set the initial conditions in radiation domination. 
While the simplest models of single field inflation give rise to nearly scale invariant adiabatic perturbations, alternative early universe models can give rise to localised features. 
In the context of inflation, these localised features will temporarily break the slow roll behaviour as the features usually come from (but not limited to) sharp features in the inflationary potential \cite{Starobinsky:1992ts, Chen:2016zuu, Romano:2014kla}.
Perhaps the more interesting set of models to test using our approach are those of bouncing or cyclic universes.
It is possible that cosine modes in a pre-bounce era source sine modes in the post-bounce era. 
Thus any signs of the sine mode in our current universe might also be a sign of a previous cycle of our universe. 
This intriguing possibility depends on how the perturbations are matched across a the bounce. 
There are various approaches to how this matching is done however most approaches depend on the underlying model that causes the bounce \cite{Bozza:2005xs, Battefeld:2005cj, Chu:2006wc, Brandenberger:2007by, Alexander:2007zm}. 

There are various natural extensions to this paper.
We have not looked at specific models in this paper however one could try to understand what is the best way to match perturbations across a bounce and what features they give rise to in the primordial power spectrum. 
Throughout this work we have assumed the cosmological parameters for the sine and cosine mode are the same. 
This does not have to be the case, as described above, and the best way to constrain the primordial and cosmological parameters together would be to do an MCMC analysis. 
We will leave a complete MCMC analysis of the adiabatic sine and cosine modes as well as the different types of isocurvature modes in addition to the cosmological parameters to future works.
In addition to scalar perturbations, one can also ask whether the most general tensor perturbations have been understood. 
Since tensor perturbations also have a second order differential equation that is the equation of motion they also must have two independent solutions and these topics are currently being explored. 

\section*{Acknowledgments}
\noindent
We would like to thank Derek Inman, Neil Turok, Pedro Ferreira, David Alonso, Emilio Bellini and Harry Desmond for useful comments and discussions. 
We thank Garrett Goon and Sadra Jazayeri for providing feedback on this manuscript.
DK acknowledges financial support from ERC Grant No: 693024. 
We would like to thank Dick Bond for a constructive argument in the CITA cosmology group meeting that led to the development of this project.
P.~D.~M.\ and DK would like to thank CITA for their hospitality while this work was being completed. P.~D.~M.\ acknowledges support from Senior Kavli
Institute Fellowships at the University of Cambridge and the Netherlands organization for scientific
research (NWO) VIDI grant (dossier 639.042.730).

\bibliography{all_active}

\end{document}